\documentclass[prl,aps,twocolumn]{revtex4}
\usepackage{graphicx} 
\usepackage[usenames]{color}
\usepackage{amsmath,amssymb}
\usepackage{gensymb}
\usepackage{natbib}
\usepackage{xcolor}
\usepackage{subfigure}
\def\strutdepth{\dp\strutbox}
\def\nw#1{\strut\vadjust{\kern-\strutdepth\vtop to0pt{\vss\hbox to\hsize
{\hskip\hsize\hskip5pt$\leftarrow$\hss\strut}}}{\em #1}}
\usepackage{rotating}

\newcommand{\invt}{\ensuremath{\tau_0^{-1}}}

\begin{document}

\title{Emergence and persistence of flow inhomogeneities \\ in the yielding and fluidization of dense soft solids}
\author{Vishwas V. Vasisht}
\author{Gabrielle Roberts}
\author{Emanuela Del Gado}
\affiliation{Department  of  Physics,  Institute  for Soft  Matter  Synthesis and Metrology,
Georgetown  University,  37th and O Streets,  N.W., Washington,  D.C. 20057,  USA}
\date{\today}
\begin{abstract}
The response to shear of the dense soft solids features a stress overshoot and a persistent shear banding before reaching a homogeneously flowing state. In 3D large scale simulations we analyze the time required for the onset of homogeneous flow, the normal stresses and structural signatures at different shear rates and in different flow geometries, finding that the stress overshoot, the shear band formation and its persistence are controlled by the presence of overconstrained microscopic domains in the initially solid samples. Being able to identify such domains in our model by prevalently icosahedrally packed regions, we show that they allow for stress accumulation during the stress overshoot and that their structural reorganization controls the emergence and the persistence of the shear banding. 
\end{abstract}

\maketitle

Dense soft solids such as emulsions, foams or colloidal pastes, yield and eventually flow under an imposed shear deformation, a feature important for materials from wet cement to food, paint or pharmaceutical products. Controlling the flow properties upon yielding is tough, since the evolution towards the steady-state is often accompanied by strong spatial inhomogeneities, where only part of the material flows while the rest stays jammed. Such phenomenon is called shear banding and has been known for many decades to geologists and engineers, but the question of what favors the stress accumulation and the persistence of flow inhomogeneities upon yielding is fundamentally unanswered. 

At a fixed imposed shear rate $\dot{\gamma}$, the shear stress increases with the strain $\gamma$ and often overshoots, followed by a decay towards a steady-state value that depends only on the shear rate. A consistent theoretical description of the fundamental physical mechanisms that control yielding in such materials is challenging and is a topic of intense debate \cite{Barrat_Review_2011, jaiswal_PRL_2016, biroli2016breakdown, leishangthem_Ncomm_2017, parisi2017shear}. Equally challenging is to control the flow properties upon yielding because of the emergence of a pronounced and persistent shear banding \cite{wright2002physics}. With the advancement in rheological tools and computational abilities, there has been a surge in the number of studies analyzing the flow inhomogeneities in yielding soft solids \cite{fielding_RPP_2014, Bonn_RMP_2017, Divoux_ARFM_2016, shrivastav_PRE_2016, shrivastav_JOR_2016}, but their origin and persistence remain elusive. 

Flow induced structuring or ordering transitions may underlie the development of bands flowing at different rates in various complex fluids, and may be affected by effective interactions between molecules, particles or droplets \cite{becu_PRL_2006, olmsted_rheoacta_2008, besseling_PRL_2010, shereda_PRL_2010, chaudhuri_pre_2012, irani_PRL_2014, chikkadi_PRL_2014, nicolas_PRL_2016}. For dense soft solids, the flow inhomogeneities delaying the fluidization are thought to emerge from the relaxation of stress heterogeneities elastically stored in the material during the stress overshoot \cite{adams2009nonmonotonic, adams2011transient, moorcroft_PRL_2011, fielding_RPP_2014, Divoux_ARFM_2016, hinkle_JOR_2016}. Simulations of jammed suspensions of Brownian particles or molecular glasses indicate indeed that the age of the sample or the cooling rate used to prepare the glass (and likely to control frozen-in stress heterogeneities invariably produced during the solidification of an amorphous structure) have an impact on the stress overshoot \cite{utz_PRL_2000, varnik_JCP_2004}. Mesoscopic models that account for elasto-plasticity of amorphous solids predict shear banding as the result of the competition between the aging of the initially solid sample and its rejuvenation under deformation \cite{Bonn_RMP_2017, martens_SM_2012, Puosi_SM_2015, patinet_PRL_2016}, again highlighting the role of mechanical heterogeneities that are, on the other hand, hard to pin down in experiments, theory and simulations. 


Here we show that the tendency to develop flow inhomogeneities upon yielding and the persistence of the shear banding are controlled by the amount of locally overconstrained, stiffer domains in the initially solid samples, using large scale 3$D$ computer simulations of a model dense non-Brownian suspension with purely repulsive interactions that undergoes yielding and fluidization. The overconstrained domains, identified in our model solids by a predominantly dodecahedral geometry of the Voronoi volumes associated to a prevalently icosahedral particle packing, allow for stress accumulation during the stress overshoot. They organize in space into a non-flowing band as shear stresses relax after the overshoot. The progressive and slow erosion of the non-flowing band helps reorganize the mechanical constraints in the material and eventually leads to the complete fluidization. 

Our model soft solid is a non-Brownian suspension of volume fraction $\phi
\approx 70\%$, consisting of $10^{5}$ (97556, unless otherwise specified) polydisperse particles, whose repulsive effective interactions are mimicked via a truncated and shifted Lennard-Jones potential: $U(r_{ij}) = 4\epsilon \left [(a_{ij}/r_{ij})^{12} -(a_{ij}/r_{ij})^{6} \right ] + \epsilon$, for $r_{ij} \le 2^{1/6} a_{ij}$, else $U(r_{ij})=0$ \cite{WCA_JCP_1971}, where $a_{ij} = (a_i + a_j)/2$ with $a_i$ and $a_j$ respectively the diameters of particles $i$ and $j$, and $\epsilon$ is the unit energy. The diameters of the particles are drawn from a Gaussian distribution with variance of 10\%, whose mean is used as unit length $a$. Albeit simple, this type of model has been successfully used for numerical simulations of sheared soft solids and proven to capture several aspects of their fundamental physics \cite{varnik_PRL_2003, irani_PRL_2014, shrivastav_JOR_2016, vasisht_arxiv_2016}. 

We prepare the initial samples from a high temperature dense liquid cooled down to low temperature, using a NVT Molecular Dynamics protocol with a cooling rate $\Gamma$ varying from $5\cdot10^{-2}$ to $5\cdot10^{-6} \epsilon/(k_{B} \tau_{0})$ (where $\tau_{0} = a \sqrt{m/\epsilon}$ is the MD time unit and for the lowest $\Gamma$ we perform $10^{9}$ MD steps). Each sample is subsequently brought to the closest energy minimum and to $k_{B}T/\epsilon \simeq 0$ via energy minimization. We use linear oscillatory rheology tests to verify that all numerical samples are initially in a solid state, well beyond the jamming point (see also SI \cite{suppl}). The samples are subjected to a finite shear rate $\dot{\gamma}$ using Lees-Edwards boundary conditions (LEBC) and, independently, by confining them between two walls, one of which is moving at a velocity determined by the chosen rate (wall based shearing protocol WB). For all data, $\dot\gamma$ is expressed in units of $\invt$. 

For the LEBC shearing protocol we solve the damped equation of motion $m \frac{d^2\vec{r}_i}{dt^2} = -\zeta_{LEBC} \left (\frac{d\vec{r}_i}{dt} - \dot{\gamma} z_i \vec{e}_x \right ) -\triangledown_{\vec{r}_i} U $ where $m$ is the particle mass and $\zeta_{LEBC}$ is the damping coefficient (we use $m/\zeta_{LEBC} = 2.0 \tau_{0}$ which guarantees that the inertial effects are minimal \cite{vasisht_arxiv_2016}). For the WB protocol, two walls confine the samples along the direction $\hat{z}$, at a relative distance $L_{z}$: one wall moves at a velocity $\vec{v} = v_{x}^{wall} \hat{x} = \dot{\gamma}L_z \hat{x}$), while the other is kept fixed. For WB the equation of motion is $m \frac{d^2\vec{r}_i}{dt^2} = - \zeta_{WB} \sum_{j(\ne i)} \omega(r_{ij}) (\hat{r}_{ij}\cdot \vec{v}_{ij})\hat{r}_{ij} - \triangledown_{\vec{r}_i} U$, where for the damping the sum for particle $i$ extends on the neighboring particles $j$ within a cut-off distance of $2.5 a_{ij}$ (with $\omega(r_{ij})=1$)  
and $\zeta_{WB}=0.1 m/\tau_{0}$ (this choice guarantees minimal inertial effects and flow properties comparable to the ones obtained with LEBC) \cite{maloney_JPCM_2008, salerno_PRL_2012}. In all simulations the stress tensor $\sigma_{\alpha \beta}$ ($\left(\alpha, \beta\right) \in \{x,y,z\}$) is computed using the virial expression \cite{irving1950statistical}. All simulations have been performed with LAMMPS \cite{LAMMPS_JCP_1995}.

\begin{figure}[h]
	\includegraphics[width=0.45\textwidth]{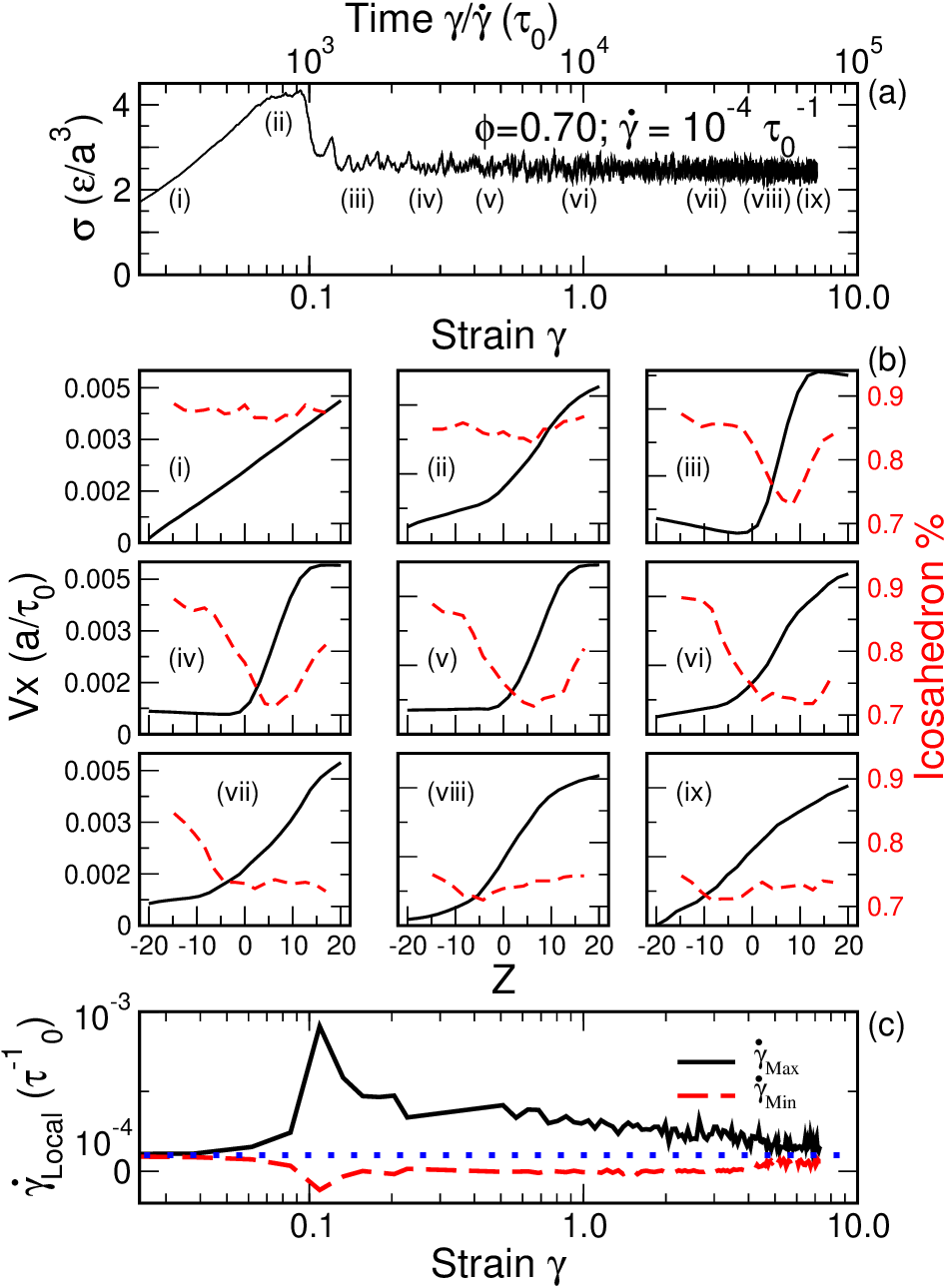}
\caption{(a) The load curve at shear rate $\dot{\gamma}=10^{-4} \invt$ for
a system prepared at $\phi=0.70$ using a cooling rate $\Gamma=5\cdot10^{-4} \epsilon/(k_{B} \tau_{0})$. The shear stress as a function of strain (and time) shows an initial elastic response, followed by an overshoot in stress, which eventually decays to reach a steady state value. (b) For various regions of the load curve, indicated by the numbers between $i$ to $ix$, we plot the velocity profile (solid line) and percentage of icosahedral packing (dashed line) as a function of the coordinate in the gradient direction. (c) The maximum (solid black line) and minimum (dashed red line) of the local shear rate computed as the numerical derivative of velocity profile, along with the applied shear rate(dotted blue line) is shown as a function of as a function of strain.}
\label{Fig1}
\end{figure}

The load curve (i.e., the shear component $\sigma$ of the stress tensor $\sigma_{\alpha \beta}$ plotted {\it vs.} the strain $\gamma$ or time $ =
\gamma/\dot{\gamma}$) is shown in Fig.\ref{Fig1} for $\dot{\gamma} = 10^{-4} \invt$ (LEBC) and features an initial elastic response and a stress overshoot at small strain (or short time). After the overshoot, the shear stress decays towards a steady-state value. At various times along the load curve (Fig.\ref{Fig1} (a)), we reconstruct the velocity profile $\langle v_{x}\rangle (z)$ averaged over a strain window of $\approx 2\%$ (Fig.\ref{Fig1} (b)), by dividing the sample in slices of thickness $\simeq a$ along the gradient direction $\hat{z}$ and averaging the $x$ component of the velocity over all particles (roughly 4000) contained in the same slice \cite{suppl, vasisht_arxiv_2016}. Starting from a linear velocity profile, as expected at short times (i), we detect the initiation of the shear banding close to the stress overshoot (ii), where the local shear rate starts to deviate from the imposed one (Fig.\ref{Fig1} (c)). By the time the stress starts decaying, part of the material is stuck in a non-flowing band: the deviation from the applied shear rate is maximum at this point (Fig.\ref{Fig1} (c)). As the stress further decays (iii) we observe a back-flow, similar to the unloading of an elastic material as proposed in \cite{adams2011transient, agimelen2013apparent, fielding_RPP_2014}. Progressive restoration of the linear velocity profile (iv)-(v) is associated to a weak but continuous decrease \cite{adams2011transient} and significant fluctuations (not shown) of the shear stress \cite{majumdar2008nonequilibrium}.

\begin{figure}[h!]
	\includegraphics[width=0.45\textwidth]{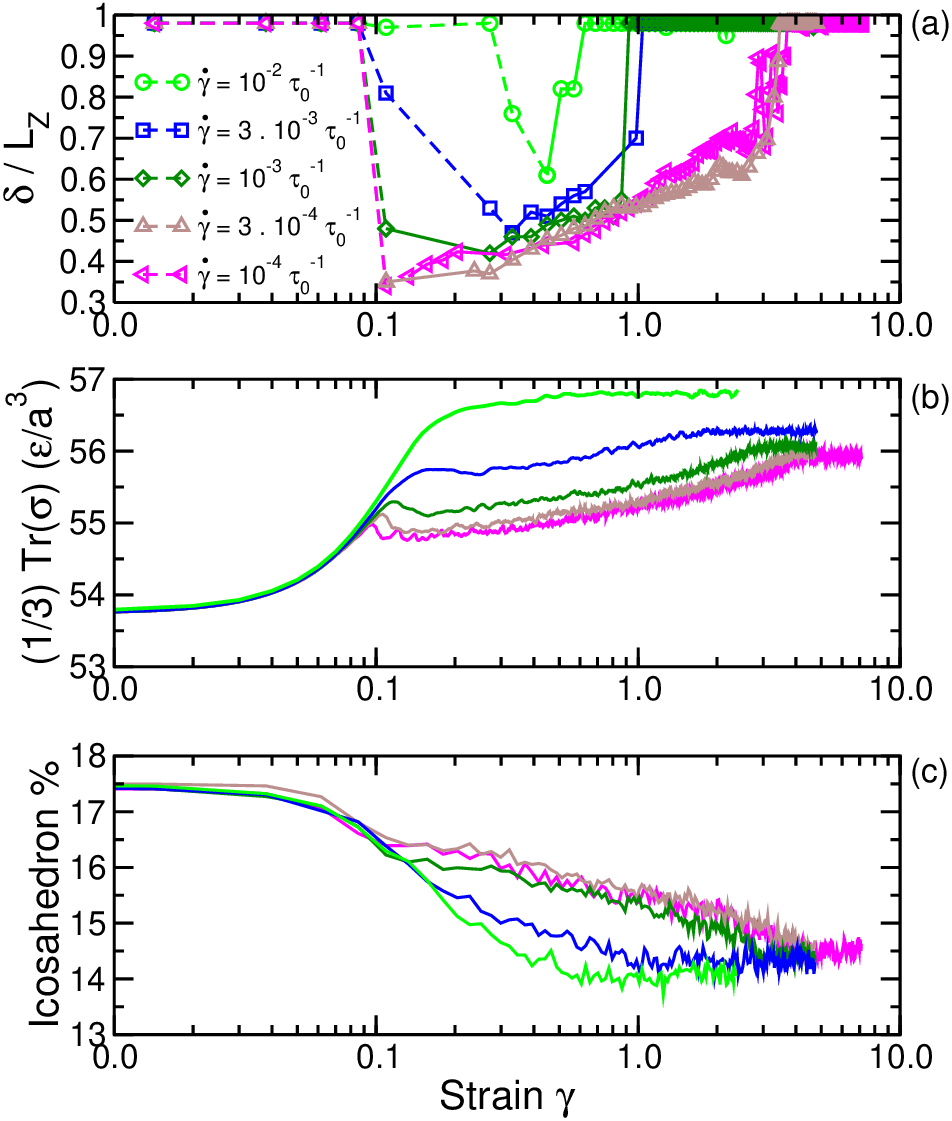}
        \vspace{5 mm}
        \caption{(a) The width of the shear band $\delta/L_{z}$ as a function of the strain $\gamma$ for different shear rates, $10^{-2} \invt \le \dot{\gamma} \le 10^{-4} \invt$, for $\phi=0.70$ and cooling rate $\Gamma=5\cdot10^{-4} \epsilon/(k_{B} \tau_{0})$. The corresponding evolution of pressure in (b) and of the percentage of icosahedral packing in (c) for the same set of shear rates. 
	}
        \label{Fig2}
\end{figure}

We monitor the fluidization through the width $\delta$ of the flowing band (measured from the velocity profiles), which evolves over time and depends on the applied shear rate
\cite{olmsted_rheoacta_2008, manning_PRE_2009, divoux_prl_2010, fielding_RPP_2014}. Fig. \ref{Fig2} (a) shows $\delta/L_{z}$ as a function of $\gamma$ (where $Lz$ is the box dimension in the gradient direction and the data refer to LEBC) starting from $1$ at small strains when the whole system is deformed homogeneously (and elastically, as indicated by the negligible dependence on $\dot{\gamma}$), dropping to a lower value that depends strongly on the rate and approaching logarithmically $1$ when the flow becomes homogeneous 
\cite{divoux_prl_2010}. By defining the fluidization time $\tau_f$ as the time required for the system to flow homogeneously after the shear start-up, the complete fluidization is clearly signaled by the evolution of the normal components of the stress (the trace of stress tensor is plotted in Fig.\ref{Fig2}(b)), indicating that normal stresses are strongly coupled to the flow inhomogeneities. The same features persist when we consider the samples sheared with the WB protocol (see also SI \cite{suppl}). 

\begin{figure}[h]
        \includegraphics[width=0.46\textwidth]{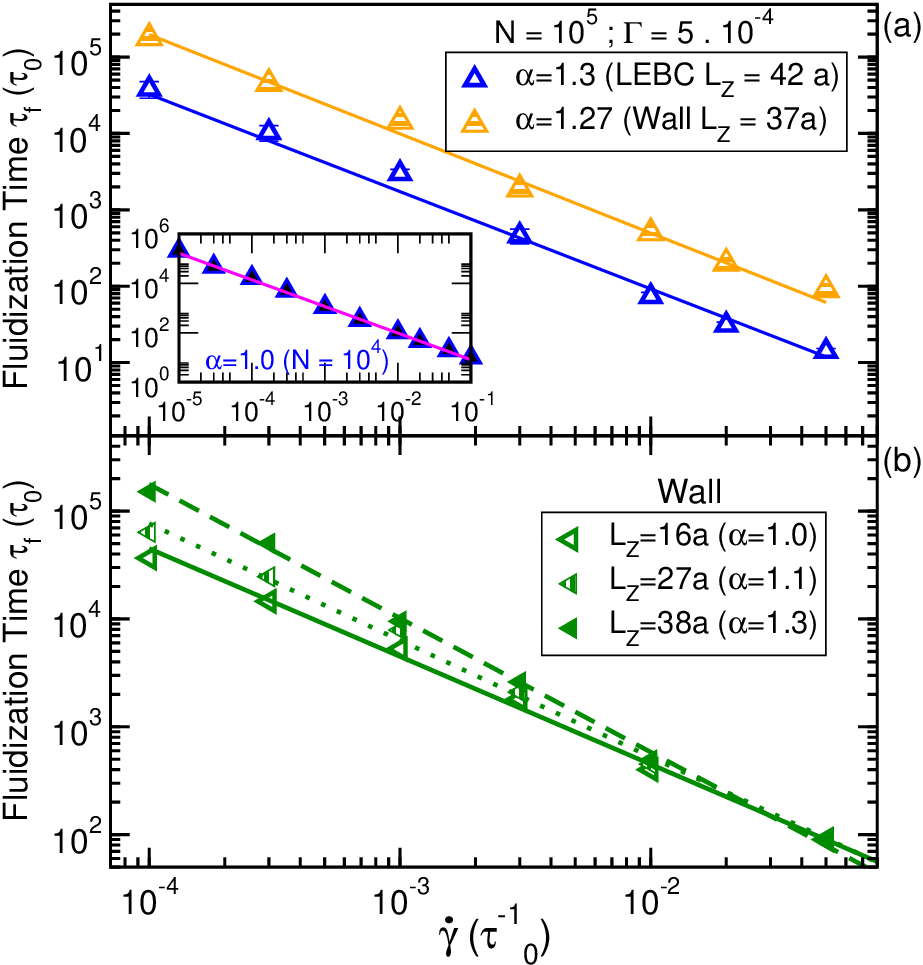}
\caption{Fluidization time $\tau_f$ as a function of shear rate for $\phi=0.70$ and initial configuration prepared at $\Gamma=5\cdot10^{-4} \epsilon/(k_{B} \tau_{0})$. In (a) main panel corresponds to $N=10^{5}$ particles, LEBC protocol (open triangle) and wall based protocol with confinement length of $37 \sigma$ (striped triangle). The error bars in the LEBC case were obtained from 3 statistically independent samples, whereas for the WB protocol they were obtained by varying the fitting range, since only one sample was available for each shear rate. The (a) inset corresponds to data at $N=10^{4}$ using LEBC protocol. In (c) we show  data for wall based protocol with different confinement lengths. }
        \label{Fig3}
\end{figure}

For relatively small system sizes ($N = 10^{4}$; $L_z = 20a$), the dependence of the fluidization time $\tau_f$ on the shear rate is $\tau_f \propto \dot{\gamma}^{-1}$ (Fig. \ref{Fig3} (a) inset), for both the LEBC and WB protocol, indicating that the time scale needed for the complete fluidization is simply set by the imposed shear rate. For large systems ($N = 10^{5}$; $L_z=42a$), instead, $\tau_f \propto \dot{\gamma}^{-\alpha}$ with $\alpha \simeq 1.3$ for both shearing protocols (Fig. \ref{Fig3} (a)). The fluidization times are always longer in the WB protocol due to the wall-bulk interface interactions, but the value of $\alpha$ is consistent with LEBC and hence likely to be dictated by bulk processes. The system size dependence of the fluidization exponent $\alpha$ is confirmed by increasing the confinement distance in the gradient direction $\hat{z}$ in the WB protocol (Fig. \ref{Fig3} (c)). The fact that $\alpha > 1.0$ for large enough system sizes indicate that the microscopic dynamical processes underlying the fluidization are not trivially slaved to the shear rate $\dot{\gamma}$, because they are spatially correlated over large distances that increase with the sample size. Such findings hint to a nucleation process or to a critical-like growth for the flow inhomogeneities, as proposed for steady state banding in complex fluids but here at play for a transient banding \cite{olmsted_rheoacta_2008}.

The non-flowing band and its persistence are not obviously associated to significant density gradients or shear induced crystallization, but we find a striking link between the persistent shear banding and a local structural signature. Given that our system is polydisperse, we construct a Voronoi tessellation to obtain the statistics of different polyhedra that corresponds to different local packing geometries and particle coordination numbers \cite{gellatly_JNCS_1982, tanemura_JCP_1983, rycroft_chaos_2009}. The analysis reveals that the time evolution of the percentage of particles associated to a dodecahedral Voronoi volume (or to a icosahedral packing) is strongly correlated to the shear banding mediated fluidization (Fig.\ref{Fig2}(c)). Furthermore, there is a strong spatial coupling between the development of the shear banding and the organization of the icosahedrally packed particle domains: in Fig.\ref{Fig1} (b) the dashed lines indicate the local icosahedral packing percentages along the gradient direction $\hat{z}$, using the same procedure employed to compute the velocity profiles, proving that, by the time the shear stresses start to relax from the overshoot, domains with mainly icosahedrally packed particles organize into the non-flowing band. 

The icosahedral particle packing points to the existence of regions where the local coordination number, and hence likely the number of mechanical constraints on a particle, is much higher than the one required by mechanical stability in isostatic conditions (i.e., $6$ in 3D) as at the onset of jamming \cite{wyart2005geometric, vinutha2016disentangling}. It has been recently noted that microscopic overconstrained domains in amorphous solids allow for local compression and tension to develop under load with no net force \cite{Lubensky_RPP_2015}: hence, under load, stress can be accumulated locally without necessarily changing the mechanical state of the material (e.g., yielding). Such feature could explain why the spatial organization and the amount of icosahedrally packed domains are coupled to the emergence and persistence of the non-flowing band. This picture is consistent with the idea that the transient banding is associated to the relaxation of the stresses stored in the dense soft solids through the overshoot\cite{fielding_RPP_2014, Puosi_SM_2015, patinet_PRL_2016}. While the persistence of icosahedral packing domains under shear and their participation to shear localization has also been noted in the context of locally preferred structures in supercooled liquids and glasses \cite{Rodney_EPJE_2011, pinney_JCP_2016,zhang2016dynamical}, we propose that they may signal overcoordinated (and hence overconstrained) regions, where stresses tend to accumulate under load. With this respect, local icosahedral packing would be akin to {\it self-stress states} discussed in \cite{Lubensky_RPP_2015}. 

\begin{figure}[h]
        \includegraphics[width=0.46\textwidth]{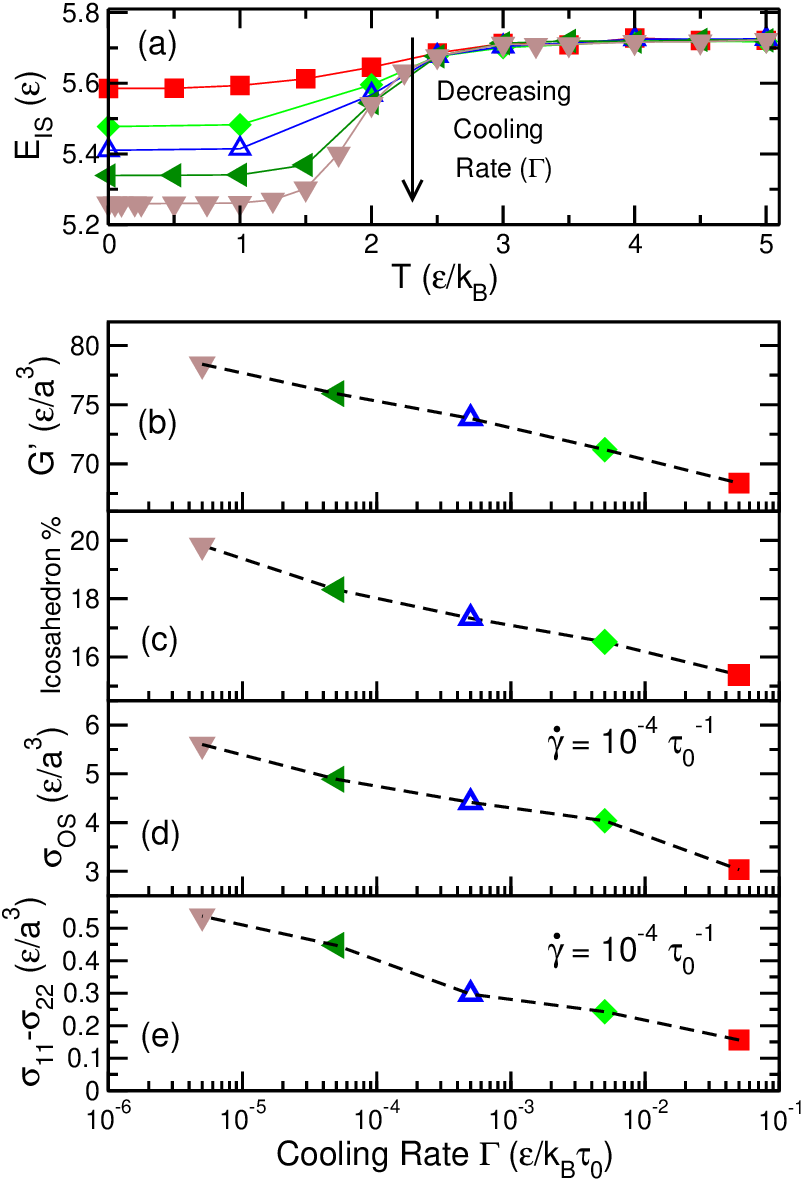}
\caption{(a) The inherent structure energy per particle as a function of the temperature for cooling rates $5\cdot10^{-2}$ (red square), $5\cdot10^{-3}$ (green diamond), $5\cdot10^{-4}$ (blue open triangle), $5\cdot10^{-5}$ (dark green triangle) and  $5\cdot10^{-6} \epsilon/(k_{B} \tau_{0})$ (brown triangle). For initial configurations (at $T = 0$ $\epsilon/k_{B}$) as a function of cooling rate $\Gamma$ we show (b) elastic modulus $G'$, (c) percentage of icosahedral packing. For initial configuration subjected to an imposed shear rate of $\dot{\gamma}=10^{-4} \invt$ we show (d) the stress overshoot and (e) the normal stress difference computed at the stress overshoot as a function of cooling rates.}
        \label{Fig4}
\end{figure}

In glassy solids and supercooled liquids whose interactions are well described by spherically symmetric potentials (of the type considered here) local icosahedral packing is known to be prevalent and corresponds to energetically favored structures that geometrically frustrate long range order \cite{steinhardt_PRL_1981}. Hence by cooling a liquid sample at different rates $\Gamma$ we can control the prevalence of icosahedral symmetry in the initial solid. Deeper local minima (or inherent structures) of the total potential energy are accessible upon decreasing $\Gamma$, as shown in (Fig. \ref{Fig4} (a)) through the inherent structure energy per particle \cite{sastry_nat_1998}. Deeper local minima also correspond to solids with higher mechanical strength as measured through the shear modulus $G'$ (Fig. \ref{Fig4} (b)) and higher percentages of local icosahedral packing (Fig. \ref{Fig4} (c)) \cite{mosayebi2012deformation, royall_PR_2015, ronceray2015favoured}. 
The logarithmic increase of the stress overshoot with decreasing $\Gamma$, as shown in Fig. \ref{Fig4} (d), here indicates that higher percentages of local icosahedral packing indeed allow for larger accumulation of stresses under deformation. The prevalence of icosahedral symmetry enhances the tendency of the material to dilate, as indicated by the first normal stress difference $\sigma_{11} - \sigma_{22}$ at the stress overshoot plotted as a function of $\Gamma$ in Fig. \ref{Fig4}(e), suggesting that icosahedral packing correspond to regions locally under compression.
Overall, all findings support the idea that the icosahedral packing particles in our model soft solids play the role of overconstrained domains that drive stress localization and eventually trigger the strain localization with the banding upon yielding. 

\begin{figure}
        \includegraphics[width=0.46\textwidth]{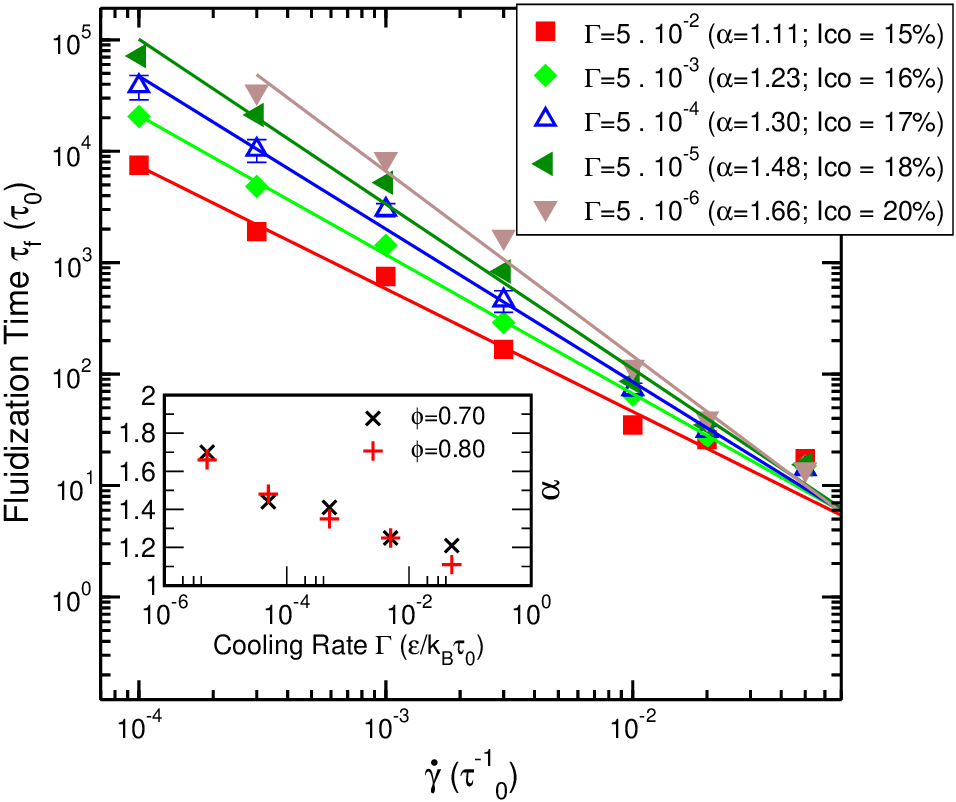}
\caption{(main panel) The fluidization time (using LEBC protocol) as a function of shear rate for initial samples obtained from different cooling rates $\Gamma$ and hence containing different percentages of icosahedral packing. (inset) The fluidization exponent as a function of cooling rates $\Gamma$ for $\phi=0.70$ (cross) and $\phi=0.80$ (plus).}
        \label{Fig5}
\end{figure}

Subjecting each of the samples prepared at different $\Gamma$ to shear deformations at different $\dot{\gamma}$, we compute the fluidization time $\tau_f$ and plot it as a function of $\dot{\gamma}$ in Fig. \ref{Fig5} (LEBC). The data show that the fluidization exponent $\alpha$ increases with decreasing $\Gamma$ and hence with the increasing icosahedral packing percentage in the initial sample. The value of $\alpha$ reach values as high as $\simeq 1.7$, indicating that for the lowest cooling rates (largest amounts of icosahedral packing) and lowest shear rates one might have to shear the samples up to $10000\%$ to get rid of flow inhomogeneities, a result that may be relevant to ultra-stable glasses \cite{singh2014ultrastable, reid2016age, berthier2017create}. The emerging picture is that the redistribution of the mechanical constraints under shear introduce a characteristic time that interfere with the imposed shear rate and strongly affects the timescale over which fluidization occurs. 

{\it Discussion.} 
We have provided the first microscopic understanding, to our knowledge, of how overconstrained domains favor stress storage (leading to a stress overshoot) in dense amorphous solids under shear deformation, by concentrating stresses in self-stress states that are mainly compressive and that self-organize into a non-flowing band in the material. As a consequence, a complete fluidization at imposed shear rate can only be attained by progressively redistributing constraints and eroding the non-flowing band. Such processes introduce a characteristic time scale that interferes with the one associated to the imposed shear rate during the start-up of the deformation, leading to a dramatic increase of the time needed to reach complete fluidization upon decreasing the shear rate. 
The true steady state behavior of the model material considered here is generic to simple yield stress fluids, well described by a Herschel-Bulkley form $\sigma-\sigma_Y \propto \dot{\gamma}^n$ (where $\sigma_Y$ is the yield stress and $n\approx 0.65$) \cite{suppl,hebraud_prl_1998}. Since the presence of overconstrained domains is also quite obvious and generic to amorphous solids, we expect that the mechanism just unveiled to control the shear banding and its persistence will have a fundamental relevance to yielding and shear localization in a variety of dense soft solids, well beyond the specifics of our numerical study. For a specific amorphous solid of interest, overconstrained domains won't be necessarily associated to dodecahedral Voronoi volumes and local icosahedral packing and may have different morphologies.

The ideas laid out here provide new input to the fundamental theoretical understanding of yielding and fluidization of amorphous soft solids, elucidating the role of mechanical heterogeneities and unraveling their spatio-temporal coupling with the imposed deformation. Recent mesoscopic theories have analyzed the role of mechanical heterogeneities in the emergence of plasticity in amorphous solids and provided a framework to understand the non-linear response in terms of the statistics and the spatio-temporal correlations of the plastic events \cite{Barrat_Review_2011}. Building on the novel insight gained here, the question to be addressed in future work is whether and how the overconstrained domains identified are indeed the microscopic fingerprints of the mechanical heterogeneities that those theories rely upon \cite{martens_SM_2012, Puosi_SM_2015, patinet_PRL_2016}. More work is also needed to quantitatively establish the connection proposed here between the overconstrained domains and the concept of self-stress states for amorphous materials \cite{Lubensky_RPP_2015}. If confirmed, such connection will shed light on the dynamical and rate-dependent implications of self-stress states, beyond their current understanding, and potentially link them to the plasticity and the plastic events statistics in a long sought-after unifying framework for amorphous solids. The implications for material science and technologies are many. Overconstrained domains could be specifically designed (through the solidification process) into smart amorphous materials to limit or enhance shear localization, depending on the specific application. They could be also used to tailor dynamics timescales that can affect material processing, with consequences for energy costs, efficiency and performances.

{\it Acknowledgments.} The authors thank Mehdi Bouzid, Thibaut Divoux, Craig Maloney, Xiaoming Mao, Kirsten Martens, Elian Masnada and Peter D. Olmsted for fruitful discussions. They acknowledge support from the National Science Foundation (Grant No. DMR-1659532), the Swiss National Science Foundation (Grant No. PP00P2 150738) and Georgetown University.
\bibliography{bibfile-Aug2017}
\clearpage
\newpage







\section{Supplementary Information}

\section{Preparation protocol}
All samples are carefully prepared using the following procedure. An initial FCC crystal containing 97556 particles at a chosen volume fraction of
$0.70$ (with $lx=ly=lz=42.1798 a$), where $a$ is average diameter of the particle, is melted at $T=5.0 \epsilon / k_B$ and the melt is equilibrated in
a NVT ensemble for around 50K molecular dynamics (MD) steps, with a timestep of $\Delta t = 0.001$. We make sure there is no signature of
crystallinity in the equilibrated sample by measuring the crystalline ordering $Q_6$ \cite{steinhardt_PRB_1983}. The equilibrated melt is subjected to
a systematic temperature quench. From the initial temperature of $T=5.0 \epsilon / k_B$ we decrease the system temperature by $\Delta T$, after which
we let the system relax at this temperature for 10K MD steps. We continue this process till $T=0.001 \epsilon/k_B$ is reached. By changing the $\Delta
T$ we control the cooling rate $\Gamma$. The samples are prepared for $\Gamma$ corresponding to $5 \cdot 10^{-2}, 5 \cdot 10^{-3}, 5 \cdot 10^{-4}, 5
\cdot 10^{-5}, 5 \cdot 10^{-6} \epsilon/(k_B \tau_0)$. Note that for the lowest $\Gamma$, we have performed total of a billion MD steps between the
initial and final temperature. After the system reaches $T=0.001 \epsilon/k_B$, we perform energy minimization using conjugate gradient method to take
the system to zero temperature limit. For $\Gamma = 5 \cdot 10^{-4} \epsilon/(k_B \tau_0) $, we have prepared three independent samples and one sample
each for the rest of the cooling rates. The initial configurations so obtained are subject to shear deformation at finite shear rate.

\section{Shearing protocols}
\subsection{Using Lees-Edwards boundary condition} 
\begin{figure}
        \includegraphics[width=0.400\textwidth]{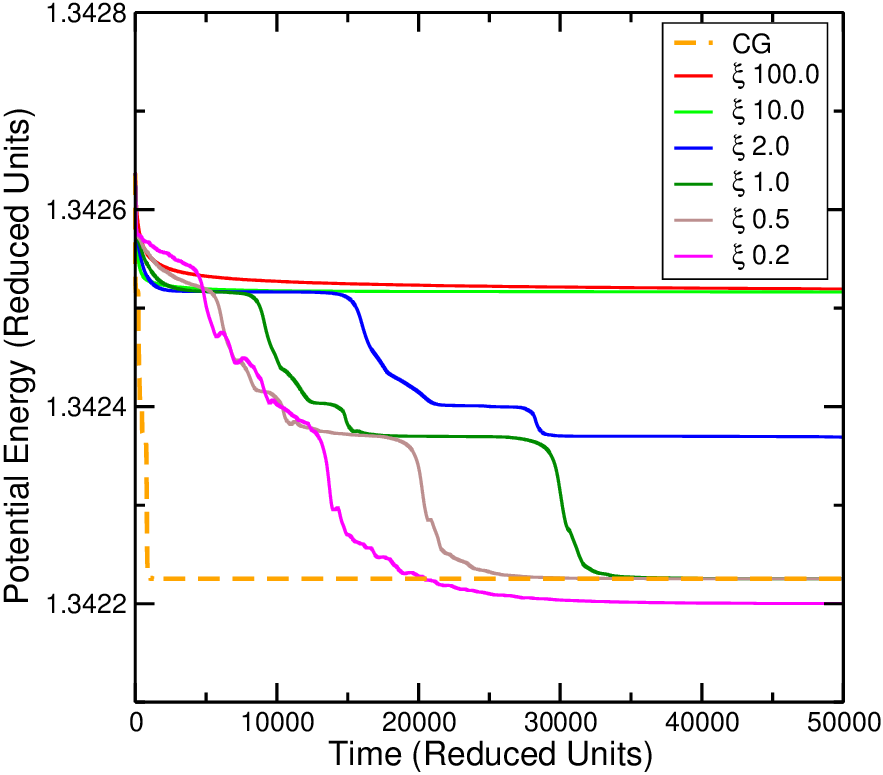}
	\caption{The potential energy as a function of time step calculated as the system relaxes after an elementary shear
strain. The orange dashed line represent the conjugate gradient minimization data and the rest of the curves corresponds to
relaxation of the system solving Langevin equation of motion for different drag coefficients.}
        \label{SupFig1}
\end{figure}

The finite shear rate rheology studies were carried out by performing shear deformation simulation
using the Lees-Edwards boundary condition and solving the Langevin equation of motion
\begin{equation}
	m \frac{d^2\vec{r}_i}{dt^2} = -\zeta_{LD} \left (\frac{d\vec{r}_i}{dt} - \dot{\gamma} y_i \vec{e}_x \right ) - \triangledown_{\vec{r}_i} U
\end{equation}
\noindent where $\dot{\gamma}$ is the applied shear rate, $m$ is the particle mass, $\zeta_{LD}$ is the damping coefficient and $m/\zeta_{LD} = 2.0$.
The damping coefficient is chosen such that, after an elementary shear deformation, the system relaxes within a reasonable amount of molecular
dynamics (MD) time steps to a potential energy (PE) which is comparable to the PE we would obtain if we performed conjugate gradient (CG)
minimization. We show in FIG. \ref{SupFig1} decay of PE as a function of time step for different damping constants and for CG minimization. The
natural time scale in the simulation is given by $\tau_0 = a \sqrt{m/\epsilon}$. We express the simulation shear rates in the units of $\invt$.\\

\subsection{Using Walls } 
\begin{figure}
        \includegraphics[width=0.400\textwidth]{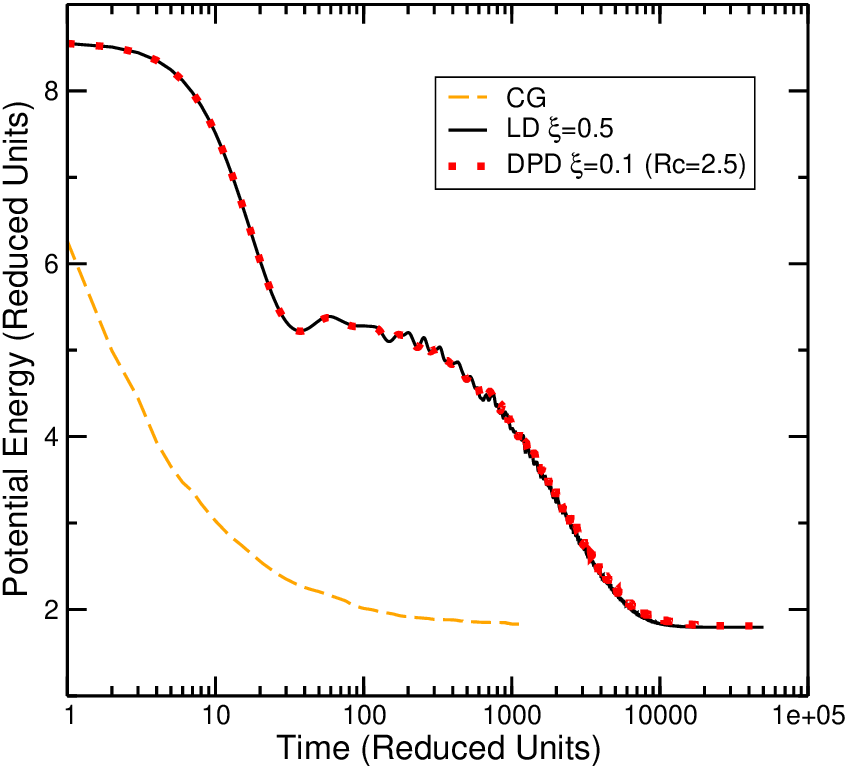}
	\caption{The potential energy as a function of time step calculated as the system relaxes after an elementary shear
strain. The orange dashed line represent the conjugate gradient minimization data. The black line corresponds to relaxation
of the system solving Langevin equation of motion for different drag coefficients and the dotted line corresponds to
dissipative dynamics with a pair-wise drag coefficient of $0.1$.}
        \label{SupFig2}
\end{figure}

To perform finite shear rate molecular simulations using walls, first we create two parallel solid walls by freezing all
the particles in the initial configuration with Y coordinates $Y_{bottom}\le -18.5899$ and $Y_{top} \ge 18.5899$ and then apply a shear strain by
moving the top wall at a velocity determined by the chosen shear rate ($U_x^{wall} = \dot{\gamma}*L_z$), while the bottom wall stay fixed.
Simultaneously we solve the dissipative equation of motion
\begin{equation}
	m \frac{d^2\vec{r}_i}{dt^2} = - \zeta_{DPD} \sum_{j(\ne i)} \omega(r_{ij}) (\hat{r}_{ij}.v_{ij})\hat{r}_{ij} - \triangledown_{\vec{r}_i} U
\end{equation}
\noindent where $\omega(r_{ij})$ is the weighting factor which is chosen as 1, the damping coefficient $\zeta_{DPD}$ is chosen to
be 0.1 and the pair-wise drag is computed within a cut-off distance of $2.5 a_{ij}$. The $\dot\gamma$ is expressed in
units of $\invt$, where $\tau_0 = a \sqrt{m/\epsilon}$ is the time unit. The damping coefficient $\zeta_{DPD}$ and the cut-off
distance is chosen such that the dissipation in the system is similar to the LD simulations. In FIG. \ref{SupFig2}, we show
the decay of PE as a function of time step for LD and DPD simulations.

\section{Structural analysis}
\begin{figure}
        \includegraphics[width=0.400\textwidth]{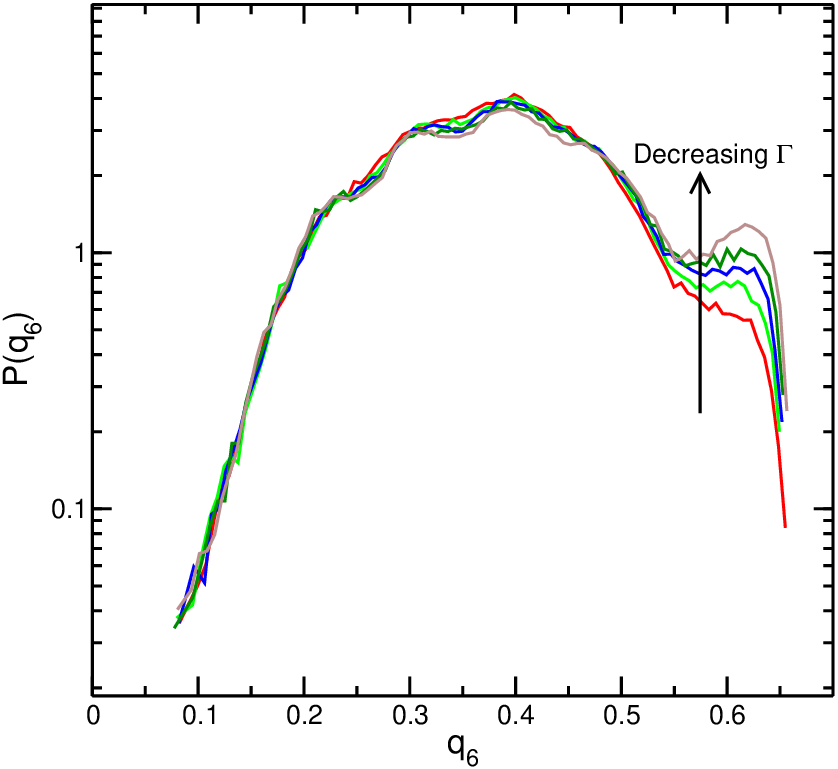}
	\caption{Distribution of orientational order parameter $q_6$ computed for initial configurations obtained from
different cooling rates.}
        \label{SupFig3}
\end{figure}

In order to compute the local structural ordering we construct radical Voronoi tessellation in 3D using the {\it Voro++} open source software library \cite{rycroft_chaos_2009} and compute the nearest neighbours to a reference particle. This method is suited for our system which is inherently polydisperse. Using the nearest neighbour information which we compute the local orientation order $q_6$ \cite{steinhardt_PRB_1983}. In the FIG. \ref{SupFig3} we show $q_6$ distribution for initial configurations obtained from different cooling rates $\Gamma$. In the $q_6$ distribution a bimodal feature appears at $q_6 \approx 0.6$ as we decreased $\Gamma$. Analyzing the Voronoi polyhedron, we obtain the information regarding the Voronoi polyhedron (faces, edges and vertices) due to the nearest neighbours. We find that the particles having a Voronoi dodecahedron (12 faces, 30 edges and 20 vertices) would predominantly contribute to the peak around $q_6 \approx 0.6$. This would correspond to an Icosahedron neighbouring environment (neighbouring atoms bonds together to form Icosahedron).

\section{Virial stress tensor}
The virial stress tensor $\sigma_{\alpha \beta}$ is computed as 
\begin{equation}
	\sigma_{\alpha \beta} = \frac{1}{V} \sum_{i} \sum_{j>i} r_{ij\alpha} f_{ij\beta}
\end{equation}
\noindent where V is the volume of the system, i and j are particle indices and $\alpha, \beta \in$ $\{X,Y,Z\}$. The distance between particle $i$
and $j$ is represented by $r_{ij}$ and force on the particle $i$ due to $j$ is $f_{ij}$. In the shearing convention following in our work (flow
direction : X, vorticity direction : Y, gradient direction : Z), the shear stress is $\sigma_{XY}$ and viral pressure is computed as $\frac{1}{3}
\left(\sigma_{XX}+\sigma_{YY}+\sigma_{ZZ} \right )$. The first and second normal stress difference would be $\sigma_{11} = \sigma_{XX} - \sigma_{YY}$
and $\sigma_{22} = \sigma_{YY} - \sigma_{ZZ}$ respectively.

\section{Linear viscoelastic response of the initial configurations}
\begin{figure}
        \includegraphics[width=0.400\textwidth]{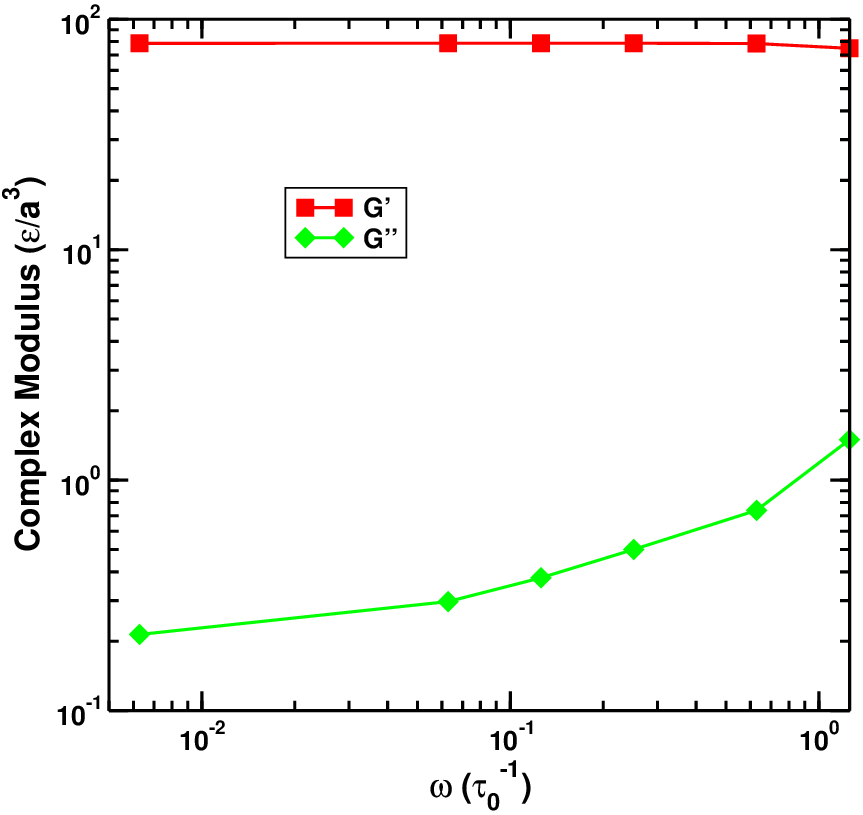}
	\caption{The elastic and plastic moduli computed as a function of frequency for an initial configuration prepared from
cooling rate $\Gamma=5 \cdot 10^{-5}$.}
        \label{SupFig6}
\end{figure}

In order to compute the complex modulus for initial configurations, we perform shear simulations under oscillatory conditions. By applying a shear
strain $\gamma$ following the equation $\gamma(t) = \gamma_0 sin(\omega t)$, for a strain amplitude of $\gamma_0 = 1\%$, we monitor the stress
response for varying frequencies $\omega$. We monitor the energy and pressure evolution with the oscillatory shear cycles and once the system reaches
a saturation in these quantities with the cycles, we extract viscoelastic coefficient using

\begin{eqnarray}
G'(\omega, \gamma_0) = \frac{\omega}{\gamma_0 \pi} \int_{t_0}^{t_0 + 2\pi/\omega} \sigma_{xy}(t) sin(\omega t) dt,\\
G''(\omega, \gamma_0) = \frac{\omega}{\gamma_0 \pi} \int_{t_0}^{t_0 + 2\pi/\omega} \sigma_{xy}(t) cos(\omega t) dt
\end{eqnarray}
\noindent In FIG. \ref{SupFig6} we show the $G'$ and $G''$ as function of $\omega$ for the initial configuration obtained from slowest cooling rate.

\begin{figure}
        \includegraphics[width=0.400\textwidth]{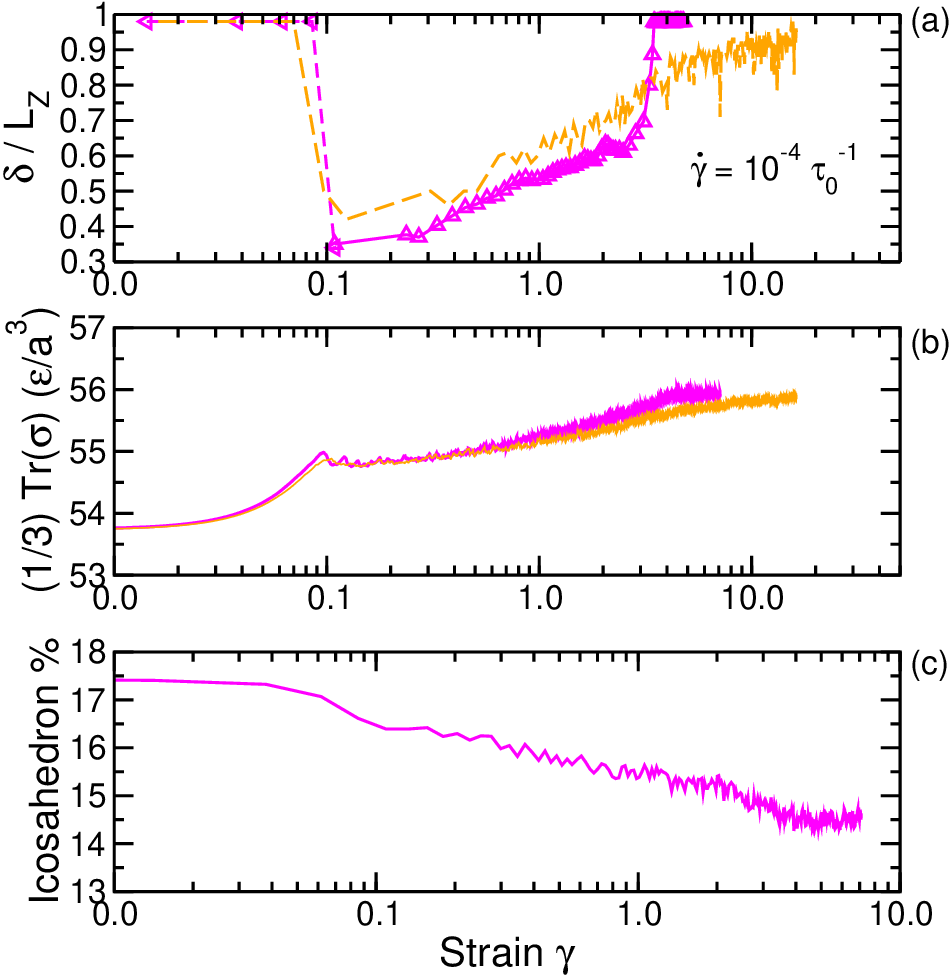}
	\caption{The width of the shear band $\delta/L_Z$, the pressure and the percentage of icosahedral packing as a function of strain, at
$\dot{\gamma}=10^{-4} \invt$, for LEBC based and wall based protocol}
        \label{SupFig11}
\end{figure}

\section{Wall and LEBC comparison}
In Fig. \ref{SupFig11}, we show the comparison in evolution of width of shear band, pressure and icosahedron percentage in the system at
$\dot{\gamma}=10^{-4} \invt$ for LEBC protocol and wall based protocol.

\section{Cooling rate and Steady state}
\begin{figure*}[h]
        \vspace{5 mm}
        \centering
        \begin{tabular}{c}
                \subfigure[]{\includegraphics[width=0.3\textwidth]{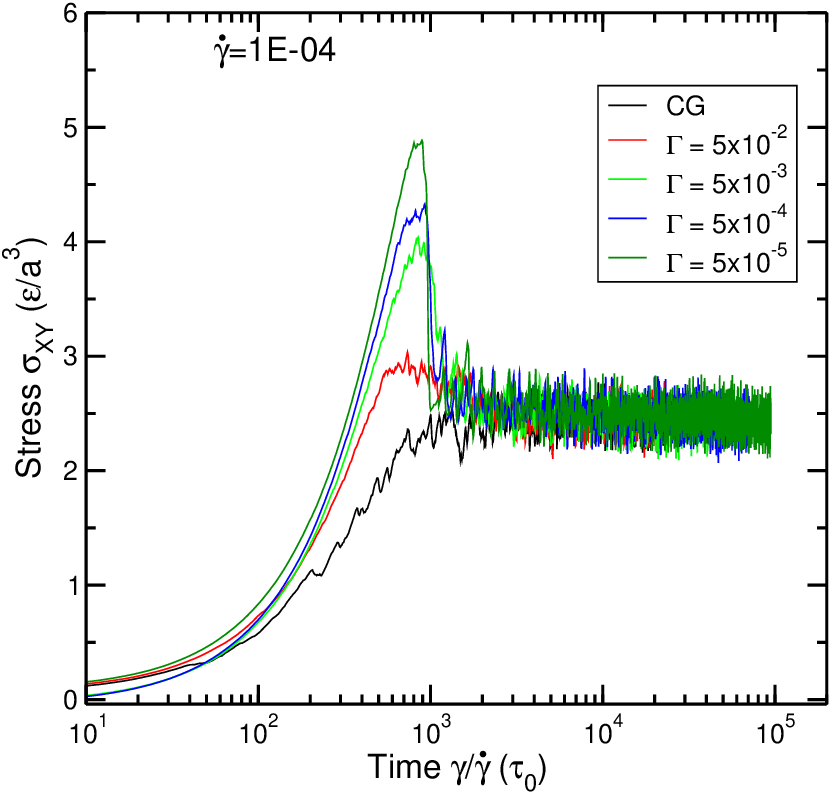}}
		\hspace{5 mm}
                \subfigure[]{\includegraphics[width=0.305\textwidth]{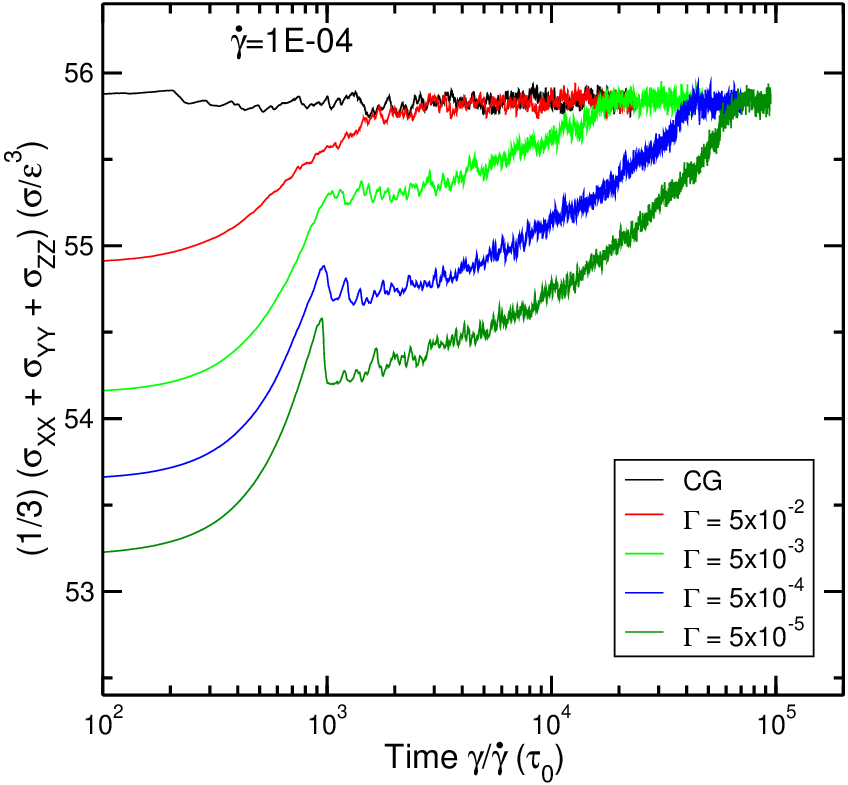}}
		\hspace{5 mm}
                \subfigure[]{\includegraphics[width=0.3\textwidth]{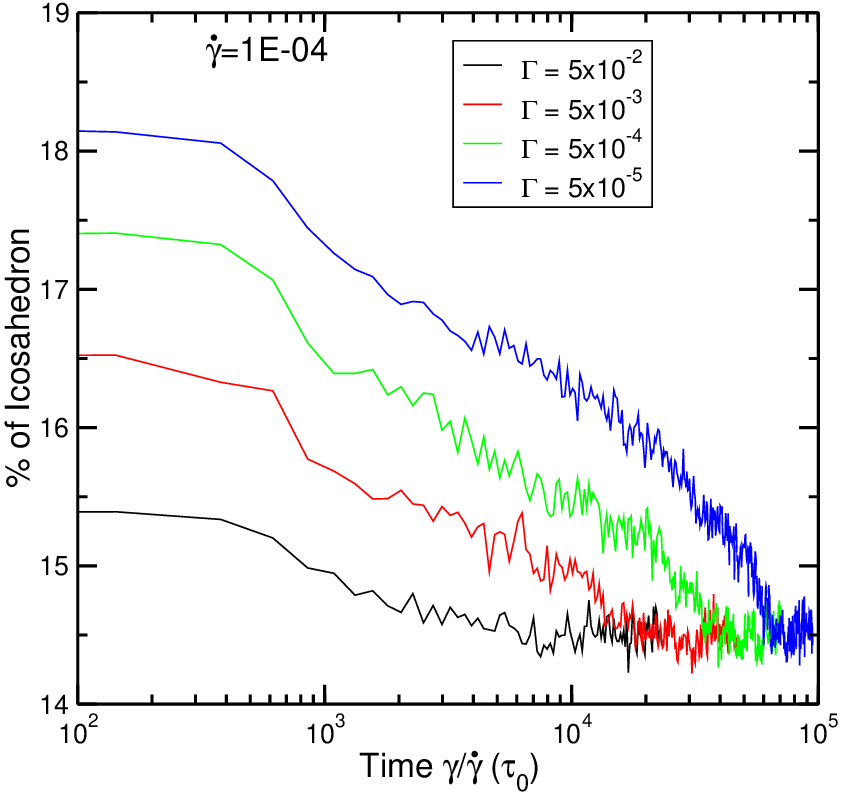}}
        \end{tabular}
        \caption{Evolution of (a) Stress, (b) Pressure and (c) Icosahedron percentage in the system with strain is shown for
samples prepared at different cooling rates and sheared at $\dot{\gamma}=10^{-4}\invt$ using LEBC protocol.}
        \label{SupFig9}
\end{figure*}

\begin{figure*}[h]
       \vspace{5 mm}
        \centering
        \begin{tabular}{c}
                \subfigure[]{\includegraphics[width=0.3\textwidth]{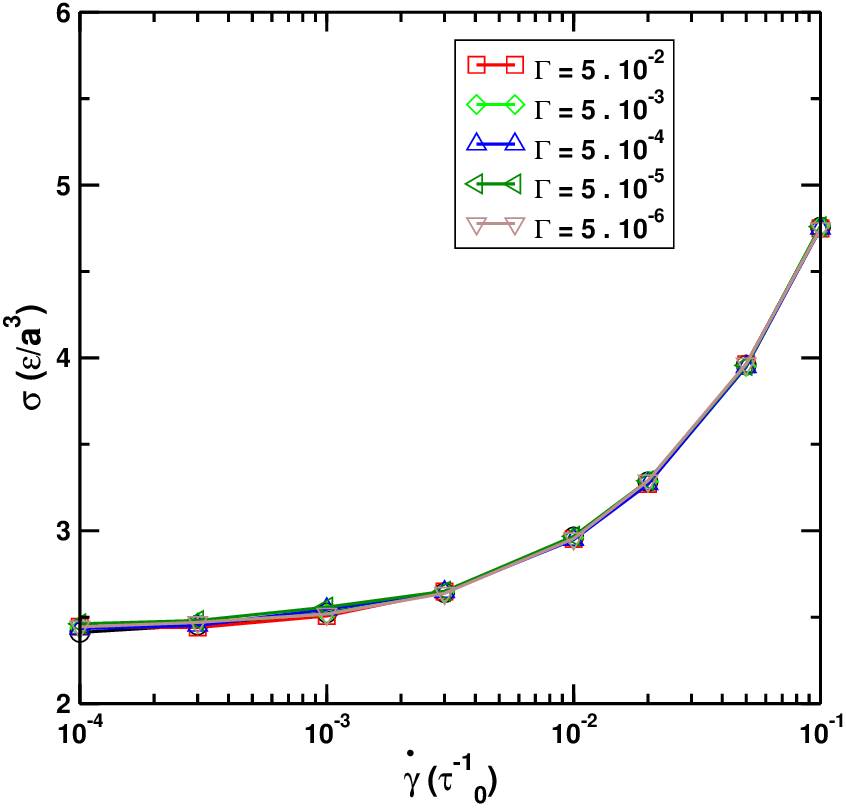}}
		\hspace{5 mm}
                \subfigure[]{\includegraphics[width=0.3\textwidth]{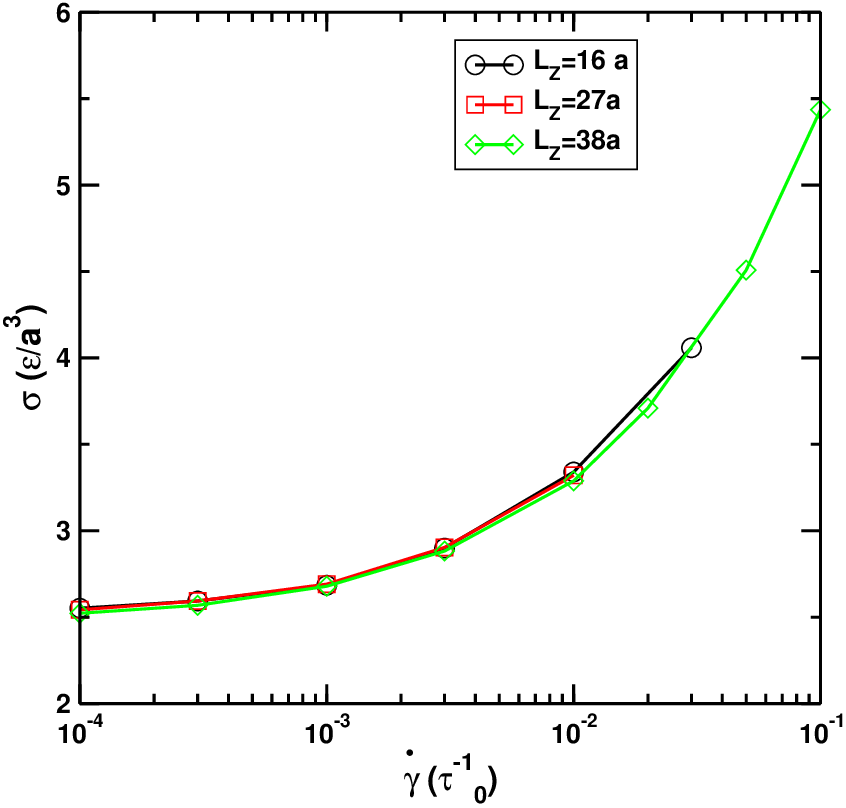}}
        \end{tabular}
	\caption{Steady state flow curve, (a) from LEBC protocol, for different initial samples prepared for a wide range of cooling rates, (b)from
wall based protocol for different confinement length (where the initial sample was prepared at $\Gamma = 5 \cdot 10^{-4} \epsilon/(k_B \tau_0)$)}
        \label{SupFig10}
\end{figure*}

As the system approaches the true steady state (determined from the fluidisation time, where we find a homogeneous flow behaviour), we do not find the dependent of sample preparation. In FIG.
\ref{SupFig9} we show stress, pressure and the percentage of Icosahedron as a function of strain for samples prepared at different cooling rates. The
steady state flow curve obtained for different cooling protocol (in LEBC protocol) and in case of wall based protocol is shown in Fig. \ref{SupFig10}.
\clearpage
\newpage

\end{document}